# Winter Noctilucent Clouds Following Sudden Stratospheric Warming: First Observations

Oleg S. Ugolnikov

Space Research Institute, Russian Academy of Sciences, Moscow, Russia

E-mail: ougolnikov@gmail.com

**Abstract**
Mesospheric structures identical to summer noctilucent clouds were observed during the nights of December 17-19, 2024 in Siberian Russia. Basing on the available photo data, the mean altitude of the clouds 70.1 ± 1.5 km was measured by umbral colorimetric method. This coincided spatially and temporary with deep temperature minimum below 160K in mesosphere, followed the polar vortex displacement and warming of stratosphere below the clouds. The satellite data on temperature and water vapor is used to study the nature of this unexpected event.

**1. Introduction**

Noctilucent clouds (NLC) are the well-known summer event observed in polar and mid-latitude mesosphere (Gadsden and Schröder, 1989). Clouds consist of water ice (Hervig et al., 2001), low temperatures are required for particle nucleation. Basing on the empirical equations of Murphy and Koop (2005) for the mean NLC altitude 82 km and typical summer $H_2O$ abundance 5 ppm, the frost point temperature is as low as 147 K. It can be reached during the summer, the clouds are mostly observed in June and July in the latitude range 50-60°, where the nights are quite dark. However, lidar and satellite techniques allow studying NLC in the daytime and circumpolar summer conditions.

Multi-year statistics by SHIAMACHY instrument onboard ENVISAT satellite (Robert et al., 2009; Köhnke et al., 2018) shows significant poleward increase of clouds occurrence rate. Polar mesospheric clouds (PMC), as they are called in this case, are nearly permanent around the summer pole. However, they are detected in the interval from –30 to +70 days relative the solstice, remaining the summer event. Observations of NLC in late May and late August are rare (Dalin et al., 2020).

Deep temperature minimum in upper mesosphere during the summer is related to the propagation and breaking of internal gravity waves causing the circulation from summer to winter pole, upwelling motion and adiabatic cooling of the air in the summer mesosphere (Andrews et al., 1987). In mid-latitudes, the mean temperature at 80-85 km reaches the minimum in the beginning of July but remains slightly above the ice frost level. However, gravity waves lead to short-scale variations with amplitude of several dozens of degrees, local minimal temperatures can be below the freezing point. It is the reason of visible wave-like structure of noctilucent clouds (Jensen and Thomas, 1994; Rapp et al., 2002; Dalin et al., 2004; Ugolnikov, 2023a) that is also traced by lidars (Gerrard et al., 2004). Planetary waves also influence the occurrence of NLC, the most remarkable is 5-day period (Merkel et al., 2003).

Meridian flow reaches the winter polar latitudes, where the usual picture is opposite: poleward and downwelling motion leads to mesosphere heating. Winter temperatures can be up to 100K warmer than during the summer, no ice particles can be produced in these conditions. This "normal" mesospheric winter regime is maintained by polar stratospheric vortex, westward gravity waves are propagating upwards (Holton, 1982).

Polar stratospheric vortex can be split or displaced as the result of interaction with planetary waves. In this case the wind direction is changed from eastward to westward, the temperature in the stratosphere increases by the several dozens of degrees. This phenomenon called "sudden stratospheric warming" (SSW) was numerically described by Matsuno (1971). Inversion of zonal wind in the stratosphere changes the sign of phase velocity of gravity waves those can reach the mesosphere (Chandran et al., 2014). The poleward meridian flow stops or turns equatorward, downwelling air motion in mesosphere can be also inverted. As the result, adiabatic cooling starts, temperature drop can reach 50K in high latitudes (Liu and Roble, 2002). The picture of air circulation becomes generally similar to the summer conditions. It affects the opposite summer polar region, reducing the mean radius of summer mesospheric ice particles (Karlsson et al., 2007).

However, even during the major SSW, mesosphere temperature remained significantly above the frost level. Zonal-mean model temperatures beyond 70°N (Liu and Roble, 2002; Siskind et al., 2010; Chandran et al., 2014; Athreyas et al., 2022) were about 200 K, CMAM model (Shepherd et al., 2014), satellite measurements (de Wit et al., 2015; Zülicke et al., 2018), and OH Meinel band analysis (Walterscheid et al., 2000) give a little bit lower temperatures, 180-190 K. Mid-latitude mesosphere during SSW is warmer, minimal temperatures were registered by lidar at the altitudes 60-70 km, being not less than 210 K (Hauchecorne and Chanin, 1983).

No mesospheric ice particles or noctilucent clouds observations were reported during SSW events until the present time. Structures similar to NLC were observed during the Antarctic winter in June 1985 (Griffiths and Shanklin, 1987). There is no data on the possibility of SSW in Southern hemisphere during that time. Estimation of the altitude of the clouds gave the value 97 km, where the ice frost temperature is very low, leading the authors to the summary of non-ice nature of the observed structures.

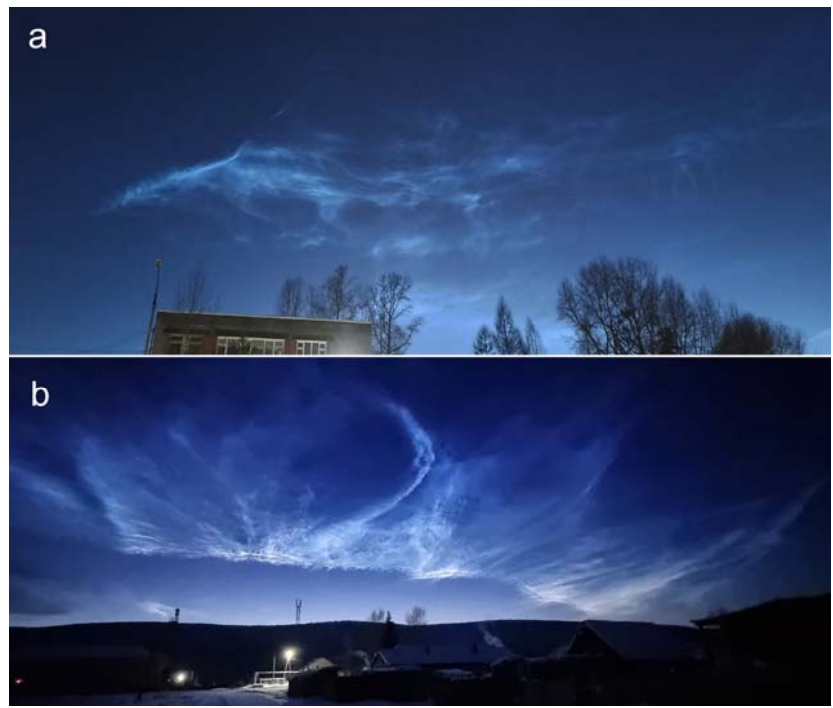

*Figure 1. Images of winter noctilucent clouds: evening twilight of December 17 (09h31m UT, 58°N, 103°E, image by Elena V. Stukalova, a) and morning twilight of December 18 (00h23m UT, 56°N, 106°E, image by Lyubov P. Yakovleva, b).*

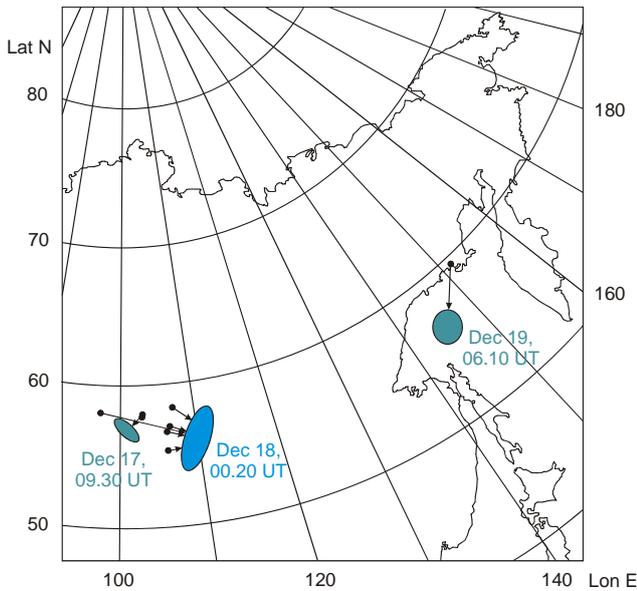
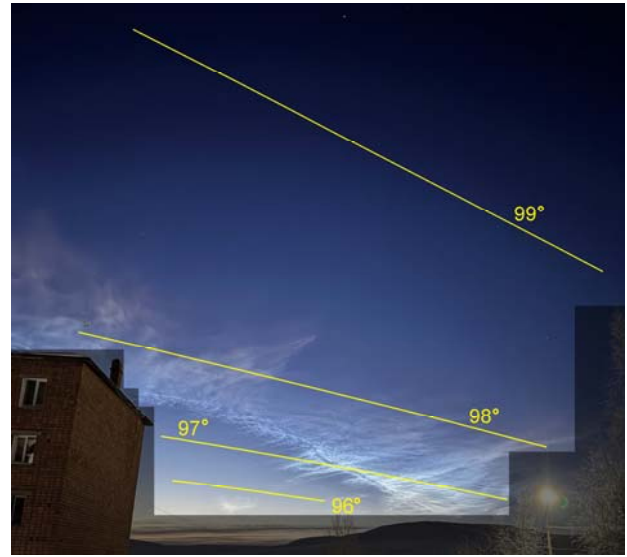

*Figure 2. Map of WNLC observations and approximate positions (evening clouds are marked dark green, morning clouds are marked blue). Mean observation times are noted.*

*Figure 3. Winter noctilucent clouds leaving the shadow of the Earth (December 18, 00h11m UT, 56.8°N, 105.7°E), image by Daria A. Sablina. Color changing effect is used for altitude definition. The values of local solar zenith angle for the altitude 70 km are shown.*

## 2. Clouds observations and altitude estimation

In the evening and the morning twilight of the night of December 17-18, 2024, bright clouds visually identical to NLC were observed in Siberian regions of Russia. They had followed by the clouds in the evening of December 19 in another location at 45 degrees to the east. Figure 1 shows the images of the clouds appeared at the dusk and at the dawn. Observation times generally differ by 2 days. The spatial area of clouds visibility was also wide – from the vicinity of Krasnoyarsk to Magadan, the map is plotted in Figure 2. This shows that the clouds were hardly the result of local atmospheric disturbance like impact of meteoric body or rocket launch.

Unfortunately, no observatories or automatic cameras were working in the locations where the clouds were visible, and the skies were clear. Winter noctilucent clouds (WNLC) were fixed by a number of occasional witnesses. There were no images taken in the same moments of time from different locations, and the stars were not seen in the most part of photographs. This did not allow running the triangulation procedure of altitude estimation, which requires precise coordinate analysis of each frame.

However, height of noctilucent cloud can be measured by single-point RGB-imaging, if it is done during the twilight stage when the cloud is immersing into or leaving the shadow of the Earth and its lower atmospheric layers. The technique is described numerically by Ugolnikov (2023b) and based on color indexes analysis. Immersing into the shadow, the cloud is first getting bluer as it is illuminated through the ozone layer in the stratosphere and then is getting red as the Sun sets beyond the troposphere. This stage was imaged in the morning of December 18 in Ust-Kut (56.8°N, 105.7°E, see Figure 3), a number of star positions was used to fix the axis direction and frame parameters.

Evolution of WNLC color due to changing conditions of solar illumination is seen in the image. Lower parts of cloud correspond to local solar zenith angle $z_L$ (visible from the cloud) about 96.5°, in this case the solar radiation trajectory lies above the dense parts of ozone layer. Blue color due to $O_3$ Chappuis absorption is seen for $z_L \sim 97°$, red color is for $z_L > 97.5°$. We can note that reddening effect starts at lower solar zenith angles than it is observed for "classical" NLC in summer (more than 98°, Ugolnikov, 2023b). It can be interpreted as lower elevation of WNLC compared with NLC. Narrow range of scattering angles does not allow fixing the mean particle size, but the altitude can be estimated.

Following Ugolnikov (2023b), we run the sky background subtraction procedure described in (Ugolnikov et al., 2021) and compare the color indexes $I_G/I_B$ and $I_R/I_B$ with model dependencies on local solar zenith angle $z_L$ for different cloud altitudes. The model uses EOS Aura/MLS satellite data on temperature and $O_3$ (NASA GES DISC, 2025), OMPS data on stratospheric aerosol (NASA GES DISC, 2025) and $NO_2$ (Yang, 2017), refraction and solar disk size with the effect of darkening to the edge. Although it is the single-scattering model, NLC heights estimated by this technique are in good agreement with triangulation altitudes (Ugolnikov et al., 2025). Influence of multiple scattering or scattering of limb radiation by NLC particles can lead to overestimation of umbral altitude of about 1 km (Ugolnikov, 2024).

WNLC intensities and colors were averaged in the circles with radius 1°. The result of comparison is shown in Figure 4. Despite the restricted data volume and low single measurement accuracy, the procedure described in (Ugolnikov, 2023b) gives the value of mean WNLC altitude: 70.1 ± 1.5 km. It is more than 10 km below the typical heights of summer NLC. However, this altitude refers to mesosphere, and these clouds can be actually considered as winter polar mesospheric or winter noctilucent clouds. Physical conditions in winter mesosphere lead to occurrence of this unusual phenomenon are discussed in the following chapter of this work.

## 3. SSW and mesosphere coupling in December 2024

To study the thermal conditions of the middle atmosphere during WNLC event, we use EOS Aura/MLS satellite data (NASA GES DISC, 2025) for December 2024. Figure 5 shows nighttime temperatures around the location where WNLC were mostly noted (55°N±2°, 105°E±10°). We see the typical picture of sudden stratospheric warming (Chandran et al., 2014; Shepherd et al., 2014; de Wit et al., 2015). It had started near December 11 with increase of temperature above 20 km and depression of stratopause to 45 km. At the same time, the mesosphere started cooling. Temperature minimum appeared at the level of pressure 4.6 Pa, it corresponds to the altitude 68 km.

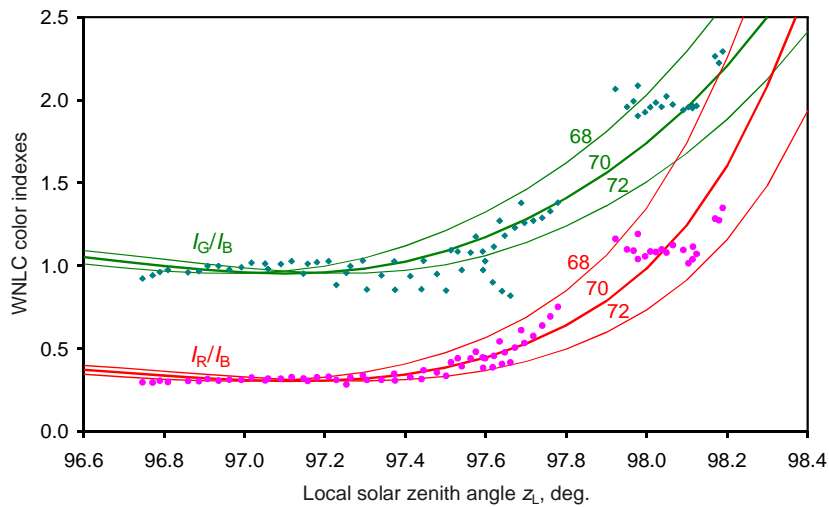

*Figure 4. Color indexes of winter noctilucent clouds depending on the local solar zenith angle, local atmospheric light extinction is reduced. Theoretical dependencies for WNLC altitudes 68, 70, and 72 km are shown for comparison.*

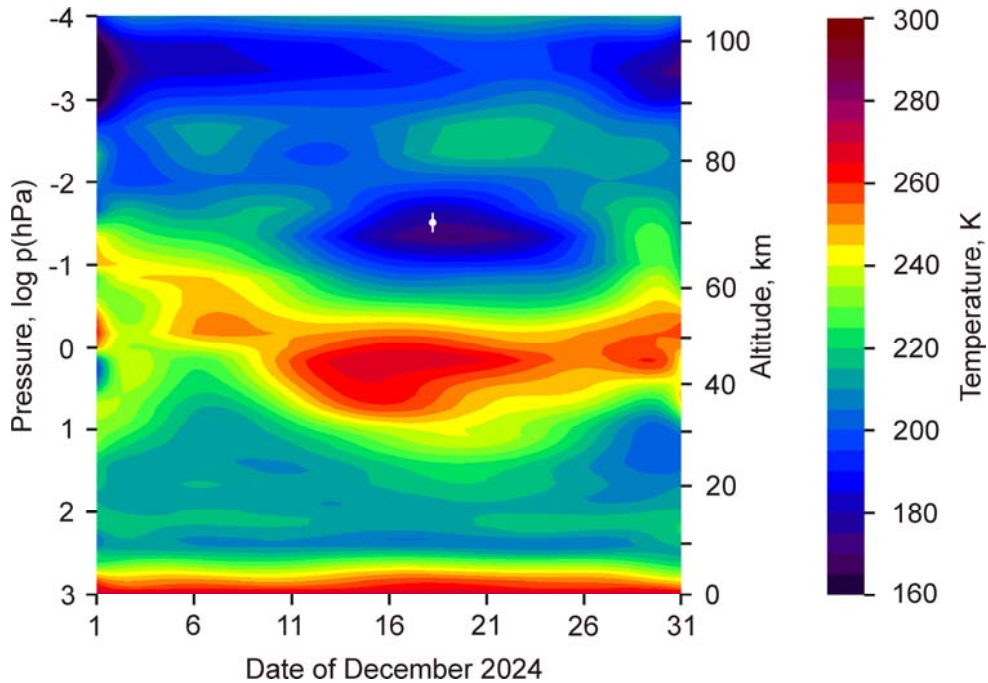

*Figure 5. EOS Aura/MLS temperatures in December 2024 averaged for 55°N±2°, 105°E±10°. Time and altitude of WNLC are shown by the dot.*

WNLC were observed after the start of maximum stage of SSW, when the temperature at this level had fallen down to 165 K according to MLS data. Height of WNLC found in previous chapter is 2 km above this level. The difference is comparable with error of measurements, we also remember that "umbral" altitude of noctilucent clouds can be overestimated for about 1 km owing to contribution of multiple scattering.

Dynamical evolution of SSW and mesosphere coupling can be seen in Figure 6, where distributions of MLS temperatures in northern latitudes are shown for the levels 147 Pa (about 44 km, maximal warming) and 4.6 Pa (68 km, deepest cooling). Average temperatures are presented for pre-SSW, beginning, maximum, final, and post-SSW stage. Mesosphere temperature minimum is shifted eastwards from the maximum of stratospheric temperatures. The same effect is seen for ascending stage of SSW in 2013 (de Wit et al., 2015).

Polar stratospheric vortex was not split or destroyed, it was displaced to the Western hemisphere during SSW, while the cold mesosphere region had shifted to the mid-latitudes above the central Siberia. WNLC were visible in the center and the east of this area. Times of WNLC observations show the probable eastward motion of the temperature minimum and clouds area. WNLC could be also observed in between but had probably missed owing to weather conditions or low population density.

Mesospheric temperature measured by MLS in the vicinity of WNLC is much lower than typical minimal temperatures during sudden warming events in stratosphere. However, it is higher than the ice frost point (about 155 K). To study the possibility of ice nucleation, we see the vertical profiles of $H_2O$ measured by MLS in the same location in December 15-18. They are plotted in Figure 7 in comparison with zonal mean profile provided by MLS for the beginning of December. Upwelling motion of mesospheric air above SSW region can lead not only to cooling, but also to water vapor transport and increase of humidity. This is actually observed in mesosphere, abundance of $H_2O$ in the layer of temperature minimum reaches almost 10 ppm at 17 and 18 of December. High density of $H_2O$ in mesosphere was already fixed during the SSW by MLS data (Manney et al., 2009). Both factors are combined to increase the probability of ice freezing.

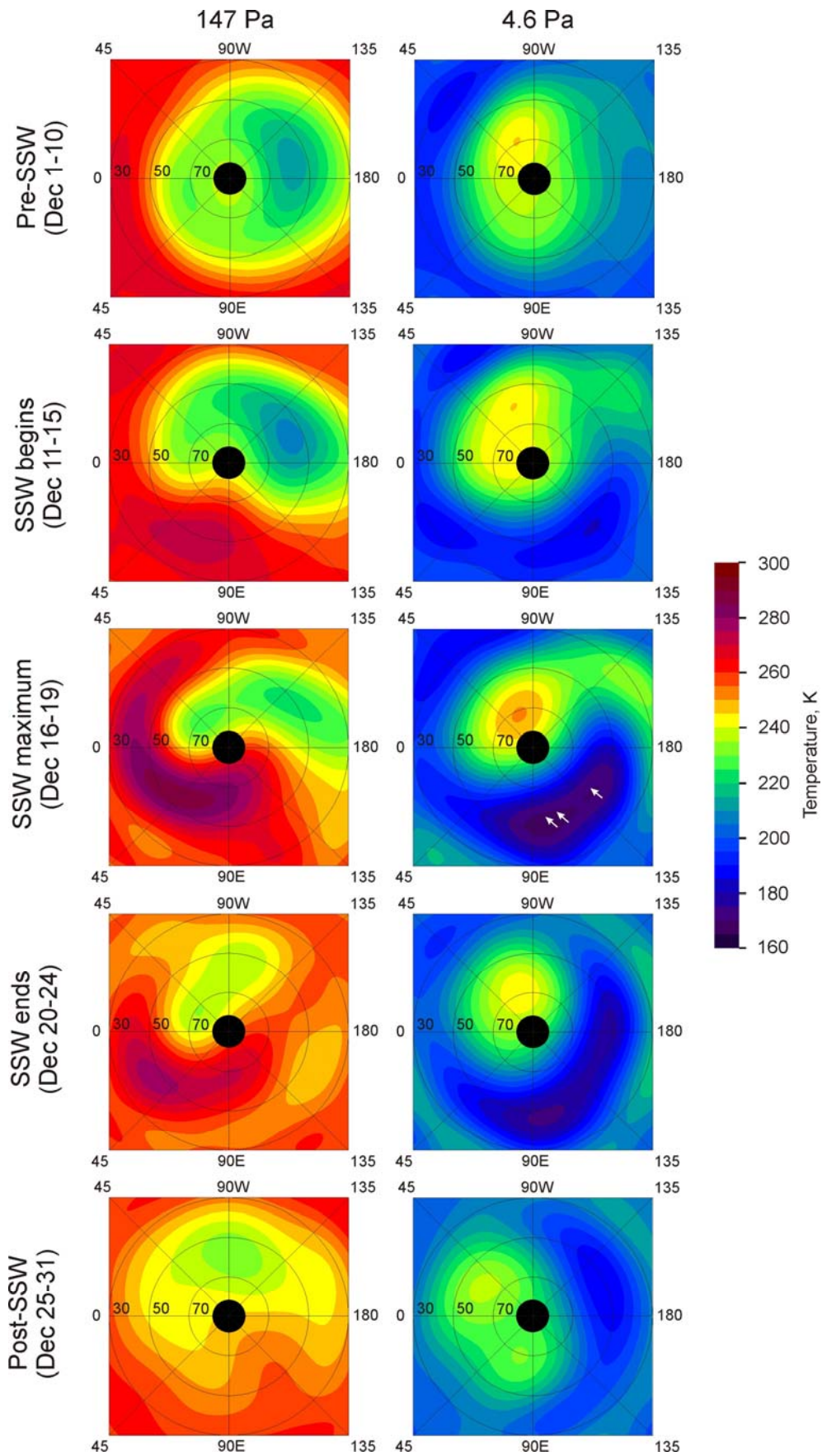

*Figure 6. EOS Aura/MLS temperatures at 147 and 4.6 Pa (approximately 44 and 68 km) averaged for different stages of stratosphere warming in December 2024. Positions of WNLC are shown by arrows in 4.6 Pa diagram for December 16-19.*

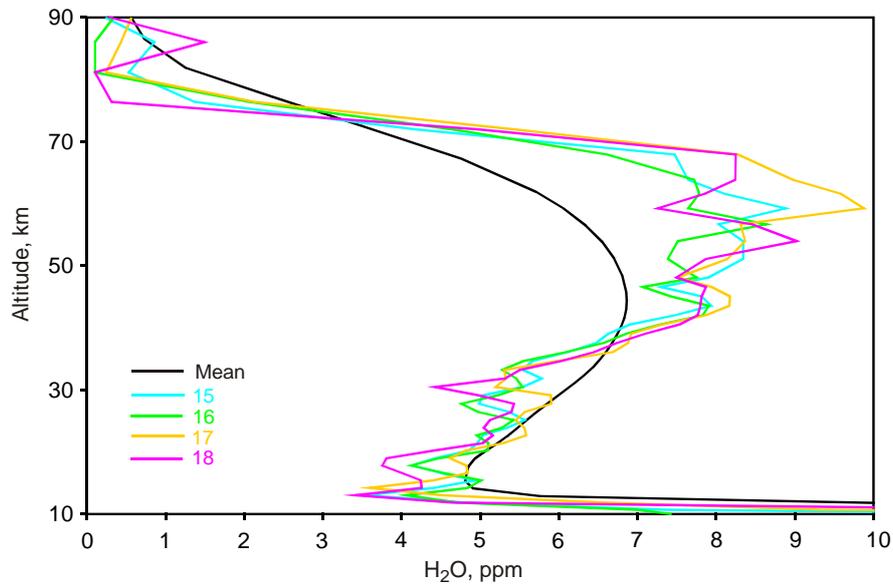

*Figure 7. EOS Aura/MLS altitude profiles of water vapor for December 15-18 2024, the same locations as in Figure 5, compared with zonal mean profile for non-SSW conditions.*

Figure 8 shows the MLS temperature profiles on the same dates and the mean profile for December 1-10, the same location area as in Figures 5 and 7. Frost temperature is calculated by equation of Murphy and Koop (2005) and $H_2O$ data of December 18, it is also shown. MLS temperatures are 10K above the frost level. The same picture is often seen in the summer mesosphere, when noctilucent clouds appear. Vertical resolution of MLS data is about 4 km, the temperature can vary at smaller scales with amplitude of dozens of degrees as the result of gravity waves propagation. The nighttime temperature profile of TIMED/SABER satellite data for December 17-18, closest to the same location is also shown in Figure 8. We see that temperature reaches frost point at the altitude 69 km. It is 1 km below the measured height of WNLC, the reason of this difference was discussed above. Since temperature variations caused by gravity waves can be short-lived, we can assume that mesosphere temperatures locally dropped even deeper during that night, starting the process of ice nucleation and occurrence of bright WNLC.

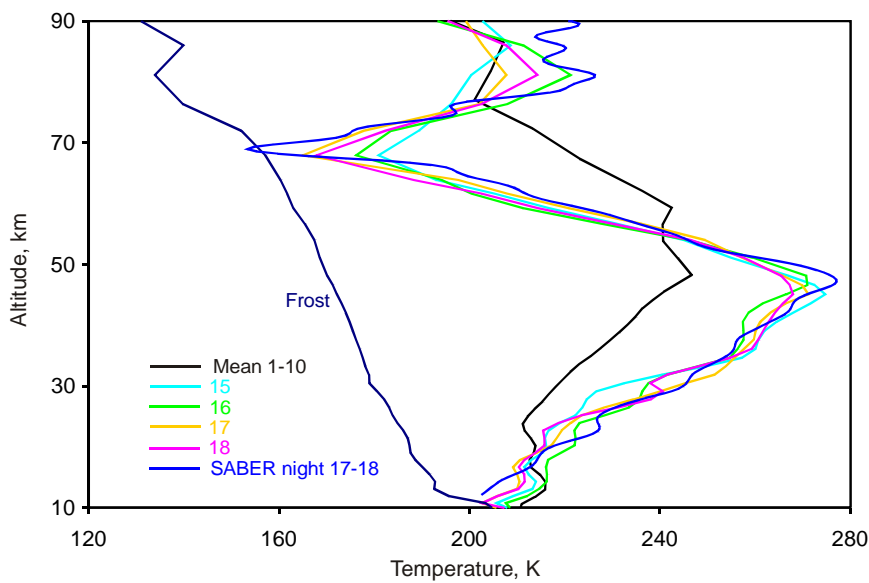

*Figure 8. EOS Aura/MLS temperature profiles for December 15-18 2024, the same locations as in Figure 5, compared with mean profile for December 1-10 and ice frost temperature. Nearby temperature scan of SABER for the night of WNLC is shown.*

## 4. Discussion and conclusion

In this paper, the phenomenon of winter noctilucent clouds at December 17-19, 2024 is discussed. Taking into account the time and place of observations, expansion of the area of WNLC visibility, and the structure of the clouds, we can be sure that they are not caused by rocket launches or other technical processes.

Meteoric origin of the clouds could be assumed. Two weeks before WNLC occurrence, small asteroid 2024 XA$_1$ (C0WEPC5) with diameter about 70 cm (MPEC 2024-X68, 2024) had entered and was probably destroyed in the atmosphere above eastern Siberia, close to the locations of WNLC observations. Impact of large body produces the dust belts in the middle atmosphere expanded along the zonal circle with latitude of entrance. This was observed in 2013 after Chelyabinsk meteorite impact (Gorkavyi et al., 2013), the dusk/dawn structures visually similar to high-altitude clouds were observed during that period. However, the dust appeared at stratospheric altitudes, about 30-40 km. Meteoric body 2024 XA$_1$ was incomparably less than Chelyabinsk meteorite, while WNLC observed now were higher and brighter than dust clouds in 2013. Just before WNLC occurrence, on December 14, meteoric shower Geminids had reached the maximum. According to International Meteor Organization (www.imo.net) reports, its activity was usual, the same as annually observed. High density of dust in mesosphere can be the secondary effect for the ice nucleation, but it can not be its primal reason.

The first key factor for understanding the nature of out-of-season noctilucent clouds is extremely low temperature at 68-70 km above the observation sites. During a few days before and after WNLC it was about 165 K according the satellite EOS Aura/MLS data. High-resolution profiles of TIMED/SABER had revealed the deep minima, where the temperature reached the ice frost level 155 K, that is 70K below the mean temperature for this latitude and season! Second key factor is the equality of the altitude of temperature minimum and the altitude of clouds measured in this paper by colorimetric analysis near the border of the shadow of the Earth.

The reason of mesosphere cooling is upwelling air motion from the stratopause during the sudden stratospheric warming started near December 11. Polar stratospheric vortex was displaced to the western longitudes, while 44 km temperature reached 300K at the opposite side from the pole. It is interesting to note that SSW events when polar vortex is displaced are more frequent during the epochs of solar activity maximum (Ageyeva et al., 2017). The question – were the unusual thermal characteristics of SSW in December 2024 related with strong solar activity – remains challenging.

It is also difficult to say whether the winter noctilucent clouds associated with sudden stratospheric warming occurred in mesosphere for the first time, and what is the mean frequency of such phenomena on Earth. Weather conditions in mid- and polar latitudes during the winter are rarely good. The clouds were observed above the wide territory, but they looked compact from each site, showing the small size of areas where temperature fell below the ice frost level. These reasons explain high probability of missing this event in the past and in the future.

It is known that occurrence of noctilucent clouds during the summer can be considered as the observational marker of climate changes in summer polar mesosphere, especially the trends of temperature and H$_2$O abundance (Lübken et al., 2018). Negative trend of mesospheric temperatures is also observed in the winter (Bremer and Berger, 2002), but it has smaller value and is basically driven by decrease of stratospheric ozone in late 20$^{th}$ century (Bremer and Peters, 2008). Due to very large difference of mean winter and ice frost temperatures, small trends are not expected to influence on the occurrence rate of WNLC. Clouds occurrence will be basically driven by planetary wave activity and thermal characteristics of SSW. According to this, we can assume that winter noctilucent clouds will remain a very rare event.

**Acknowledgments**

Author thanks Daria A. Sablina, Vadim A. Boyarinov, Maxim P. Bulgakov, Eugene V. Byankin, Sergey M. Efimenko, Stanislav A. Korotkiy, Ivan A. Neelov, Alexey V. Popov, Vladimir A. Scheglov, Elena V. Stukalova, Andrey V. Timofeev, Egor Yu. Tsimerinov, and Lyubov P. Yakovleva for their help while gaining the information and the images of winter noctilucent clouds.